\documentstyle[sprocl]{article}

\input{psfig.sty}
\input{amssymb.sty}

\def\Journal#1#2#3#4{{#1} {\bf #2}, #3 (#4)}

\def\NPPS{{\em Nucl. Phys. Proc. Suppl.}}

\def\PRD{{\em Phys. Rev.} D}

\def\EPJ{{\em Eur. Phys. J.} C}

\def\be{\begin{equation}}
\def\ee{\end{equation}}
\def\bea{\begin{eqnarray}}
\def\eea{\end{eqnarray}}
\def\eVdist{\kern-0.06667em}

\def\gev{{\,\textrm{Ge}\eVdist\textrm{V\/}}}

\catcode`\@=11 
\def\parenbar{\mathpalette\p@renb@r}
\def\p@renb@r#1#2{\vbox{%
  \ifx#1\scriptscriptstyle \dimen@.7em\dimen@ii.2em\else
  \ifx#1\scriptstyle \dimen@.8em\dimen@ii.25em\else
  \dimen@1em\dimen@ii.4em\fi\fi \offinterlineskip
  \ialign{\hfill##\hfill\cr
    \vbox{\hrule width\dimen@ii}\cr
    \noalign{\vskip-.3ex}%
    \hbox to\dimen@{$\mathchar300\hfil\mathchar301$}\cr
    \noalign{\vskip-.3ex}%
    $#1#2$\cr}}}
\catcode`\@=12 

\begin{document}
{\small
\vspace*{-1.2cm}
\begin{flushright}
BONN-HE-2000-02 \\
June 2000
\vspace{0.2cm}
\end{flushright}
}
\title{\boldmath STUDY OF DEEP INELASTIC $ep$-SCATTERING AT HIGH $Q^2$ \\ WITH ZEUS
  AT HERA}

\author{A. KAPPES\footnote{Supported by grants from the 'Bundesministerium
    f\"ur Bildung und Forschung' in Germany and the 'German-Israeli
    Foundation'.} \\ (On behalf of the ZEUS COLLABORATION)}

\address{Physikalisches Institut, Universit\"at Bonn, Nu{\ss}allee 12, 
D-53115 Bonn, Germany\\E-mail: kappes@physik.uni-bonn.de} 

\maketitle\abstracts{ Measurements of differential cross sections for deep
  inelastic $ep$ neutral-current (NC) and charged-current (CC) reactions are
  presented using data taken during the 1994--97 $e^+p$ and the 1998--99 $e^-p$
  running periods. The structure function $xF_3$ has been extracted combining NC
  $e^-p$ and $e^+p$ data sets. Both the measured cross sections and $xF_3$ are
  well described by the standard model predictions using standard PDF's.  }

\section{Introduction}
During the running periods 1994--97 ($e^+p$) and 1998--99 ($e^-p$), ZEUS has
collected data corresponding to integrated luminosities of $47\, \mbox{pb}^{-1}$
and $16\, \mbox{pb}^{-1}$, respectively. These data allow the investigation of
the high-$Q^2$ regime both in the $e^+p$ and the $e^-p$ channel. For CC the new
results from $e^- p$ are compared to previously published $e^+p$ data. For NC
new results from $e^-p$ and the extracted $xF_3$ points are shown.  For CC
$e^+p$ ($e^-p$) the kinematic range is $Q^2 > 200 \gev^2$ ($Q^2 > 1\,000
\gev^2$) and for NC $e^-p$ it is $Q^2 > 200 \gev^2$ and $ 0.0032< x < 0.65$.
$xF_3$ has been measured for $Q^2 > 3\,000 \gev^2$. 

It should be noted that the center-of-mass energy has been raised from $\sqrt{s}
= 300\, \mbox{Ge\kern-1pt V}$ in 1994--97 to $\sqrt{s} = 318\, \mbox{Ge\kern-1pt
  V}$ in 1998--99 by increasing the proton beam energy from $E_p = 820\,
\mbox{Ge\kern-1pt V}$ to $E_p = 920\, \mbox{Ge\kern-1pt V}$.

\section{Charged Current (CC)}

To leading order, the $e^\pm p \rightarrow \parenbar{\nu}X$ cross section can be
written as
\begin{equation}
  \frac{d^2 \sigma^{CC}(e^\pm p)}{dx\,dQ^2} =
  \frac{G_F^2}{2 \pi} 
  \left( \frac{M_W^2}{Q^2 + M_W^2} \right)^2
  \cdot 
  \left\{ \begin{array}{r@{\ :\ }l}
      \left(\bar{u} + \bar{c} \right) + \left(1-y\right)^2 \left(d +
        s\right) & e^+p \\
      \left(u + c \right) + \left(1-y\right)^2 \left(\bar{d} +
        \bar{s}\right) & e^-p
    \end{array} \ , \right. \label{eq:ddcc}
\end{equation}
where $G_F$ is the Fermi constant, $M_W$ the mass of the $W$ boson and
$\parenbar{u}$, $\parenbar{d}$, $\parenbar{c}$ and $\parenbar{s}$ are the quark
momentum distributions. Measuring the high-$x$ $e^+p$ cross section thus mainly
probes the $d$ valence distribution, whereas $e^-p$ mainly carries information
on the $u$ valence distribution. In contrast to $e^-p$ scattering, for $e^+p$
the valence contribution to the cross section is suppressed by the factor
$(1-y)^2$.

\begin{figure}[t]
  \begin{center}
    \begin{minipage}[c][9cm][c]{6.3cm}
      \psfig{width=6.3cm,figure=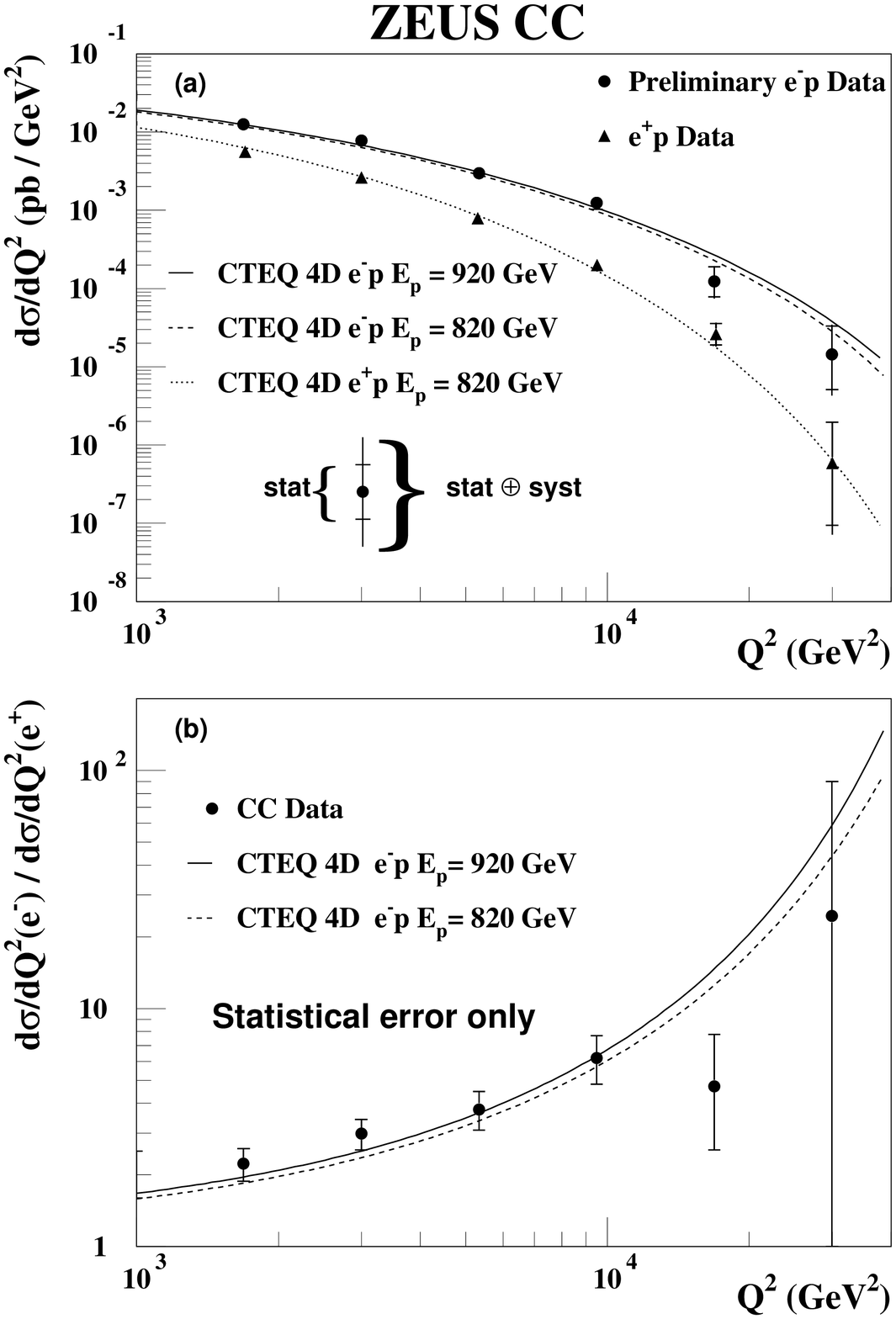}
    \end{minipage}
    \hspace*{-0.2cm}
    \begin{minipage}[c][9cm][c]{4.8cm}
      \psfig{width=5.2cm,figure=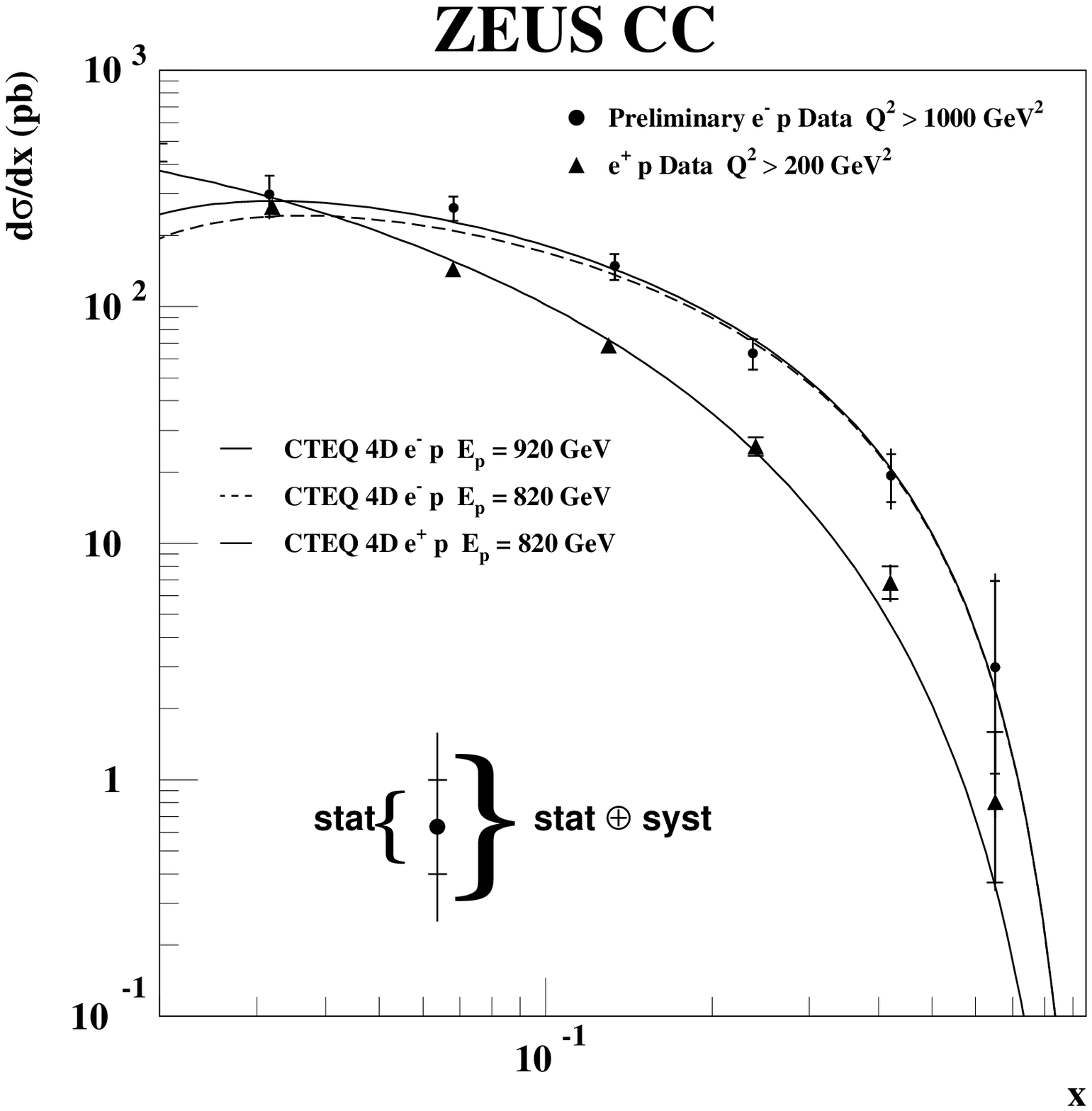}
      \begin{minipage}{5.5cm}
        \vspace*{-4.8cm}
        \hspace*{1cm}
        {\tiny \boldmath $(c)$}
        \vspace{4.4cm}
        \caption{CC cross sections for $e^+p$ and $e^-p$ data. {\bf a)}: $d\sigma / dQ^2$ vs.
          $Q^2$. {\bf b)}: Ratio between $\sigma_{e^-p}$ and $\sigma_{e^+p}$.
          {\bf c)}: $d\sigma / dx$ vs. $x$. The solid and dotted lines show the
          theoretical predictions using CTEQ4D~\protect\cite{cteq4d}. The dashed
          lines are the theoretical predictions for $e^-p$ at $E_p = 820\,
          \mbox{Ge\kern-1pt V}$.
          \label{fig:cc}}
      \end{minipage}
      \vspace*{0.5cm}
    \end{minipage}
  \end{center}
\vspace*{-1.25cm}
\end{figure}

The single differential cross sections $d\sigma / dQ^2$ for
$e^-p$~\cite{cc_tampere} and $e^+p$~\cite{cc_desy99-059} are shown as functions
of $Q^2$ in Fig.~\ref{fig:cc}a. Figure~\ref{fig:cc}b shows the ratio between the
two cross sections. The increase in the cross section by more than an order of
magnitude when switching from $e^+p$ to $e^-p$ for $Q^2 \gtrsim 10\,000 \gev^2$
is evident. It is mainly due to the higher $u$ quark content of the proton wrt.
its $d$ quark content and the additional suppression of the $d$ quark
distribution in the $e^+p$ case. Only a small fraction of the increase
originates from the increased proton beam energy. The $Q^2$ dependence of the
propagator term in Eq.~\ref{eq:ddcc} permits the determination of the $W$ mass
from a fit (not shown) to the measured $d\sigma / dQ^2$
points~\cite{cc_desy99-059}. Fixing $G_F$ to the SM value yields $M_W =
81.4^{+2.7}_{-2.6} \mathrm{(stat.)} \pm 2.0 \textrm{(syst.)}^{+3.3}_{-3.0}
\textrm{(PDF) \gev}$, which is compatible with the world average. Note that in
$ep$ scattering space-like $W$'s are exchanged whereas e.g. at Tevatron or LEP
the $W$'s are produced on-shell.

Figure~\ref{fig:cc}c shows $d\sigma / dx$ versus $x$ for $e^-p$ and $e^+p$. Note
that in the case of $e^+p$ the data are for $Q^2 > 200\, \textrm{Ge\kern-1pt
  V}^2$, whereas for $e^-p$ the region $Q^2 > 1\,000\, \mathrm{Ge\kern-1pt V}^2$
was selected. Nevertheless, for $x>0.04$ the $e^-p$ cross section exceeds the
$e^+p$ cross section. The errors on the measured cross sections are large in the
high-$x$ range and of the same order of magnitude as the uncertainties of the
theoretical predictions (not shown). Hence the data can not yet contribute to an
improvement of the uncertainties of the PDF's in this region.

\section{Neutral Current (NC)}

To leading order, the NC cross section can be written as 
\begin{equation}
  \frac{d^2 \sigma^{NC} \, (e^\pm p)} {dx\,dQ^2} =
  \frac{2 \pi \alpha^2}{x\, Q^4} \left[ Y_+ F_2^{NC} \mp Y_- 
    x F_3^{NC} -  y^2 F_L^{NC} \right] \,, \label{equ:dxq}
\end{equation}
where $Y_\pm = 1 \pm (1-y)^2$. In contrast to the charged current case, the
exchanged boson couples to all quark flavors, yielding the structure functions
\begin{equation}
  \displaystyle F_2  = x \sum_f A_f \cdot \left[ q_f + \bar{q}_f \right] 
  \hspace{1cm}
  \displaystyle xF_3 = x \sum_f B_f \cdot \left[ q_f - \bar{q}_f \right] \ ,
\end{equation}
where the sum runs over the different quark flavors $f$, $A_f$ and $B_f$ are the
electroweak coupling factors and $\parenbar{q}$ are the quark momentum
distributions. The contribution of the longitudinal structure function $F_L$ can
be neglected for the selected kinematic range. For $Q^2 \ll M_Z^2$ ($M_Z$ being
the $Z$ mass), also the contribution of $xF_3$ to the cross section can be
neglected. However, for $Q^2 \gtrsim M^2_Z$ the weak and electromagnetic forces
become comparable in size, implying a sizable contribution of $xF_3$.

\begin{figure}[t]
  \begin{center}
    \psfig{width=4.7in,height=6.9cm,figure=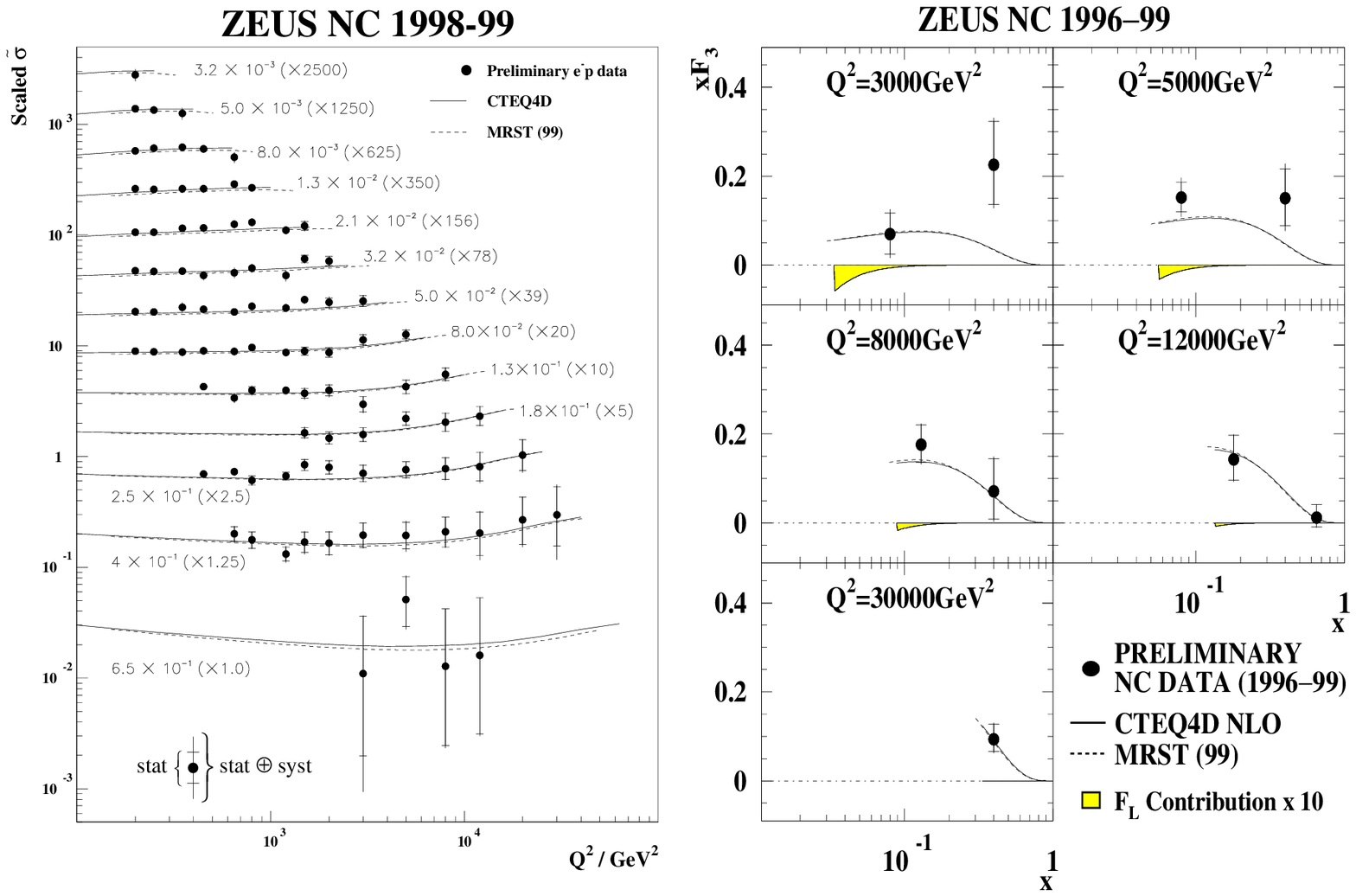}
    \vspace*{-0.7cm}
    \begin{minipage}[t]{5.8cm}
      \caption{Reduced cross section $\tilde{\sigma}$ versus $Q^2$ for fixed
        values of $x$ ($e^-p$ data). The cross sections of each $x$ bin are
        scaled by the factor in parentheses. The solid and dashed lines show the
        theoretical predictions using CTEQ4D and MRST(99)~\protect\cite{mrst99},
        respectively.
        \label{fig:rxsec}}  
    \end{minipage}
    \hfill
    \begin{minipage}[t]{5.3cm}
      \caption{$xF_3$ versus $x$ for fixed values of $Q^2$ from the combined
        $e^+p$ and $e^-p$ data sets. The solid and dashed lines show the
        theoretical predictions using CTEQ4D and MRST(99), respectively. The
        shaded band in each $Q^2$ bin indicates the $F_L$ contribution
        multiplied by a factor of 10.
        \label{fig:xF3}}
    \end{minipage}
  \end{center}
\vspace{-0.2cm}
\end{figure}

Figure~\ref{fig:rxsec} shows the $e^-p$ reduced cross section $\tilde{\sigma} =
\frac{d^2 \sigma^{NC}}{dx\,dQ^2} \cdot \frac{x\, Q^4}{2 \pi \alpha^2}
\frac{1}{Y_+}$ as a function of $Q^2$ at fixed values of $x$ ranging from $0.65$
down to $0.0032$. Both the CTEQ4D~\cite{cteq4d} and
MRST(99)~\kern-0.08em\cite{mrst99} parameterizations describe the data well over
the whole $x$ range. The cross section is dominated by $F_2$ at low $Q^2$,
where scaling violation is driven by QCD, whereas at high $Q^2$ the $xF_3$
contribution becomes significant and causes an increase in the cross section.

$xF_3$ has been determined from the difference of $e^-p$ and
$e^+p$~\cite{nc_9697} cross sections using $16\, \textrm{pb}^{-1}$ of $e^-p$
data ($\sqrt{s} = 318\, \textrm{GeV}$) and $30\, \textrm{pb}^{-1}$ of $e^+p$
data ($\sqrt{s} = 300\, \textrm{GeV}$). Figure~\ref{fig:xF3} shows the extracted
$xF_3$ versus $x$ for five values of $Q^2$ ranging from $3\,000\,
\mbox{Ge\kern-1pt V}^2$ up to $30\,000\, \mbox{Ge\kern-1pt V}^2$. Comparing the
sizes of the statistical errors to the $F_L$ contributions (corrections to the
calculated $xF_3$ arising from different $\sqrt{s}$ values of the two data
sets), calculated from CTEQ4D, justifies the assumption $F_L = 0$. The two
theoretical predictions from CTEQ4D and MRST(99) are very similar and both
describe the data well.

\section{Summary}
$47\, \textrm{pb}^{-1}$ of $e^+p$ and $16\, \textrm{pb}^{-1}$ of $e^-p$ data
have been analysed to obtain high-$Q^2$ NC and CC cross sections. The CC cross
sections $d\sigma / dQ^2$ and $d\sigma / dx$ have been measured for $Q^2 >
1\,000 \gev^2$ ($e^-p$) and $Q^2 > 200 \gev^2$ ($e^+p$), respectively. The
$e^-p$ NC cross section $d^2\sigma / dx\,dQ^2$ has been determined for $Q^2 >
200 \gev^2$ and $0.0032 < x < 0.65$. Combining the $e^+p$ and $e^-p$ data sets
made it possible to extract the structure function $xF_3$ for the first time in
the high-$Q^2$ regime, $Q^2 > 3\,000 \gev^2$.  All measured cross sections and
$xF_3$ are well described by the standard model predictions using standard
PDF's.


\section*{References}
\begin{thebibliography}{99}
\bibitem{cc_tampere} ZEUS Collab., J.~Breitweg et al., Measurement of High-$Q^2$
  Charged-Current $e^-p$ Deep Inelastic Scattering Cross Sections, contributed
  paper \#558 to ICHEP99, Tampere (1999).

\bibitem{cc_desy99-059} ZEUS Collab., J.~Breitweg et al., \Journal{\EPJ}{12}{411}{2000}.

\bibitem{cteq4d} H.-L.~Lai et al., \Journal{\PRD}{55}{1280}{1997}.

\bibitem{mrst99} A.D.~Martin et al., \Journal{\NPPS}{79}{105}{1999}.

\bibitem{nc_9697} N.~Tuning, these proceedings.

\end{thebibliography}
\end{document}